\documentclass[prb,amsmath,amssymb,shownopacs,twocolumn]{revtex4}
\usepackage{graphicx}
\usepackage{subfigure}

\newcommand{\beq}{\begin{equation}}
\newcommand{\eneq}{\end{equation}}
\newcommand{\bea}{\begin{eqnarray}}
\newcommand{\eea}{\end{eqnarray}}

\input{epsf}

\begin{document}



\title{Band Collapse and the Quantum Hall Effect in Graphene}

\author {B. Andrei Bernevig,  Taylor L. Hughes}
\affiliation{Department of Physics, Stanford University, Stanford,
CA 94305}
\author{Han-Dong Chen}
\affiliation{Department of Physics, University of Illinois at
Urbana-Champaign, 1110 W. Green St., Urbana, IL 61801}
\author{Congjun Wu}
\affiliation{Kavli Institute for Theoretical Physics, University
of California, Santa  Barbara, CA 93106}
\author{Shou-Cheng Zhang}
\affiliation{Department of Physics, Stanford University, Stanford,
CA 94305}

\begin{abstract}
\begin{center}

\parbox{14cm}{The recent Quantum Hall experiments in graphene have
confirmed the theoretically well-understood picture of the quantum
Hall (QH) conductance in fermion systems with continuum Dirac
spectrum. In this paper we take into account the lattice, and
perform an exact diagonalization of the Landau problem on the
hexagonal lattice. At very large magnetic fields the Dirac
argument fails completely and the Hall conductance, given by the
number of edge states present in the gaps of the spectrum, is
dominated by lattice effects. As the field is lowered, the
experimentally observed situation is recovered through a
phenomenon which we call band collapse. As a corollary, for low
magnetic field, graphene will exhibit two qualitatively different
QHE's: at low filling, the QHE will be dominated by the
``relativistic" Dirac spectrum and the Hall conductance will be
odd-integer; above a certain filling, the QHE will be dominated by
a non-relativistic spectrum, and the Hall conductance will span
all integers, even and odd.}

\end{center}
\end{abstract}

\maketitle

The quantum Hall effect (QHE) is one of the richest phenomena
studied in condensed matter physics. This effect is characterized
by certain conductance properties in two-dimensional samples
\emph{i.e.} the vanishing of the longitudinal conductance
$\sigma_{xx}\sim 0$ along with the onset of a quantized transverse
conductance $\sigma_{xy}= \nu \frac{ e^2}{h}.$ Recently several
experimental groups  have produced two-dimensional plane films of
graphite, commonly known as graphene sheets\cite{Novoselov2005,
YZhang2005}, which exhibit interesting QHE behavior.

Graphene has a theoretical history beginning with the study of the
band structure of this planar system in \cite{Wallace1947}. From
these humble beginnings it has gone on to be studied intensely
because of its Dirac structure. The bands can be effectively
characterized by massless $(2+1)d$ Dirac fermions
\cite{Semenoff1984}.  This continuum model of graphene has been
subsequently used to study the $(2+1)d$ parity
anomaly\cite{haldane1988} and as a model system for the relativistic
quantum Hall effect (RQHE)
\cite{Schakel1991,castroneto2006a,gusynin2005}. A quantum spin Hall
effect has also been predicted in graphene \cite{Kane2005a,
Kane2005b}, but the intrinsic spin orbit gap is probably too small
to support a measurable phase\cite{yao2006a,macdonald2006b}.

The latter studies were based on the recent experimental work done
on the QHE in graphene by two independent
groups\cite{Novoselov2005,YZhang2005}. These two groups confirm an
interesting behavior in graphene in which the transverse conductance
is quantized as an integer plus a half-integer
$\sigma_{xy}=(n+\frac{1}{2}) 4 e^2/h,$ where band and spin
degeneracies have been taken into account. Although unrelated to the
parity anomaly, this behavior of the Hall conductance was in fact
obvious in the seminal work of Jackiw and Rebbi \cite{Jackiw1976}.
On the basis of the argument for the RQHE
\cite{Schakel1991,castroneto2006a,gusynin2005} the experimental
groups conclude that this is an interesting new phenomena completely
explained by the relativistic Dirac spectrum of graphene. We want to
improve on this argument for several reasons.  For very large $B$
the lattice is expected to dominate the behavior of the Hall
conductance.  In this regime the Dirac argument cannot be valid,
since, by virtue of being a continuum argument, it ignores lattice
effects and the torus structure of the Brillouin zone. We will see
this is indeed the case, and the large-$B$ limit does not match the
Dirac argument prediction. On the other hand, in the experimental
situation the magnetic field is weak (with respect to the unit
quantum flux per plaquette) and the Dirac argument applies, it is
nonetheless desirable to have a description of the quantum Hall
effect valid for both strong and weak magnetic fields. At low
filling, we show how graphene evolves from a high-$B$ regime with
non-Dirac behavior to a low-$B$ regime with Dirac behavior through a
phenomenon we dub ``band collapse." Two adjacent bands close the gap
between them across the whole Brillouin zone and form a new band
with twice the degeneracy of each of the initial bands. The edge
structure reflects this degeneracy.

We begin with a restatement of the RQHE argument based on the
relativistic $(2+1)d$ Dirac spectrum. We then present the exact
solution of the Landau problem on the graphene lattice. The
agreement we find between numerical diagonalization and analytic
calculations done with Hatsugai's\cite{Hatsugai1993a} theoretical
framework lead us to our conclusions and illustrate the competition
between the relativistic and non-relativistic character of the band
structure of graphene in a magnetic field.

\section{The Relativistic Quantum Hall Effect in Graphene}
\begin{figure}[h]
\includegraphics[scale=0.35]{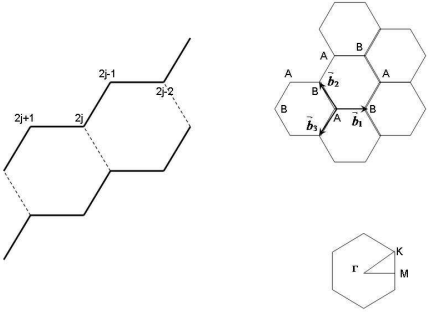}
\caption{Graphene Lattice, BZ and One dimensional lattice on which
the Harper equation is defined.} \label{graphene}
\end{figure}

We start with the tight-binding nearest neighbor Hamiltonian for the
hexagonal lattice given by Semenoff \cite{Semenoff1984}:
\begin{eqnarray}
& H = -t \sum_{\vec{A}, i} c^\dagger(\vec{A}) c(\vec{A} +\vec{b}_i)
+c^\dagger(\vec{A} +\vec{b}_i) c(\vec{A}) \nonumber \\ & + \beta
\sum_{\vec{A}} c^\dagger(\vec{A}) c(\vec{A}) -c^\dagger(\vec{A}
+\vec{b}_i) c(\vec{A} +\vec{b}_i)
\end{eqnarray} \noindent where $c(\vec{A}), c(\vec{A}
+\vec{b}_i)$ are the annihilation operators for sites on sublattice
$A$ and $B,$ and $\beta$ is an energy difference for electrons
localized on the $A$ and $B$ sublattices. We will call this term the
Semenoff mass. Graphene is effectively massless which is
approximated by taking $\beta \rightarrow 0$.
 In this limit the  band structure is gapless at two inequivalent
points $K =\frac{4 \pi}{\sqrt{3}a} (\frac{1}{2}, \frac{1}{2
\sqrt{3}}), K'=-K$, where $a$ is the nearest-neighbor lattice
constant. Around these points, the Hamiltonian is described by (in
the ideal case massless) Dirac fermions with
\cite{Semenoff1984,haldane1988}:
\begin{equation}
H_{K} = \sigma_x k_x + \sigma_y k_y ; \;\;\; H_{K'} = - \sigma_x k_x
+ \sigma_y k_y
\end{equation} \noindent which act on a two-spinor wavefunction
describing the sublattices $A$ and $B$, see Fig[\ref{graphene}].
There is also an overall 2-fold spin degeneracy which we neglect for
the remainder of the paper. Note that parity switches $A
\leftrightarrows B$ and $K\leftrightharpoons K'$ while time reversal
switches $K\leftrightharpoons K'$. The Semenoff term opens a gap of
value $m = 2 \beta/\sqrt{3} t a$ at $K$ and $-m$ at $K'=-K$ so time
reversal symmetry is preserved.

\begin{figure}[h]
\includegraphics[scale=0.35]{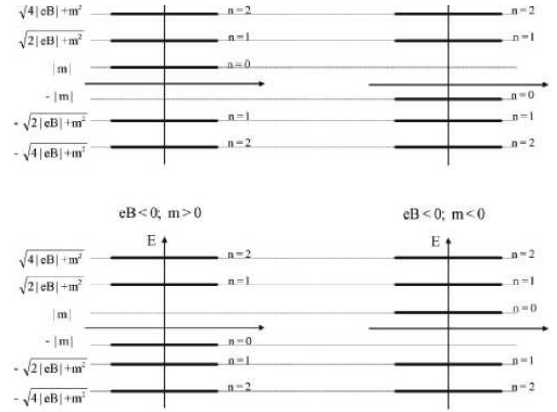}
\caption{Zero mode in the Dirac Equation.} \label{diracfermion}
\end{figure}

Now consider one Dirac fermion at the $K$-point with mass $m$ in
magnetic field $B$. The Hamiltonian is $H=\sigma_x k_x + \sigma_y
(k_y - e B x)+ m \sigma_z$. For $eB >0$, the eigenstates are
\begin{eqnarray}
  u^\pm_{k,n}=\frac{e^{iky}}{4\pi\alpha_n}
  \left(
  \begin{array}{c}
    i\sqrt{\alpha_n\pm m}~\psi_n(x-x_0^\pm(k)) \\ ~\\
    \pm\sqrt{\alpha_n\mp m}~\psi_{n-1}(x-x_0^\pm(k))
  \end{array}
  \right)
\end{eqnarray}
with
\begin{eqnarray}
 \alpha_n &=& \sqrt{2 | eB| n + m^2} \nonumber
\\  x_0^\pm &=& \frac{1}{eB} (k \pm \alpha_n)  \nonumber
\\  E^\pm &=& \pm \alpha_n \nonumber
\end{eqnarray} \noindent where $\psi_n(x)$ are harmonic oscillator eigenstates and $u^\pm$ are the eigenstates of $H_K$ with
energies $E^\pm$. Notice that all the energy levels are paired
\emph{except} the $n=0$ level. There is a common misconception that
unpaired ``zero-modes" occur only for a massless fermion but observe
that for $m>0$ we have $u^-_{k,0} =0$ while for $m<0$ we have
$u^+_{k,0}=0$, so such levels are unpaired even for non-zero mass.
In the field theory formalism, the current is defined to be $J^\mu=
-\frac{1}{2} e \gamma^\mu_{\beta \alpha} [\psi_\alpha,
\overline{\psi}_\beta]$ and is odd w.r.t. charge conjugation
symmetry. We find that
\begin{eqnarray}
 \langle 0|J^0|0\rangle =\rho = \frac{1}{2}(N_- - N_+)\frac{|eB|}{2 \pi}
\end{eqnarray} \noindent where $N_+$ and $N_-$ are the numbers of
filled positive and negative energy Landau levels (LL).  Hence the
Hall conductance is
\begin{equation}
\sigma_{xy} = \frac{1}{2}(N_{-} - N_{+} )
\end{equation} \noindent in units of $e^2/h$. Due to the unpaired level, this will be
half-integer and the position of the unpaired level depends on the
sign of $eB$ and $ m$ as in Fig[\ref{diracfermion}].

\begin{figure}[h]
\includegraphics[scale=0.35]{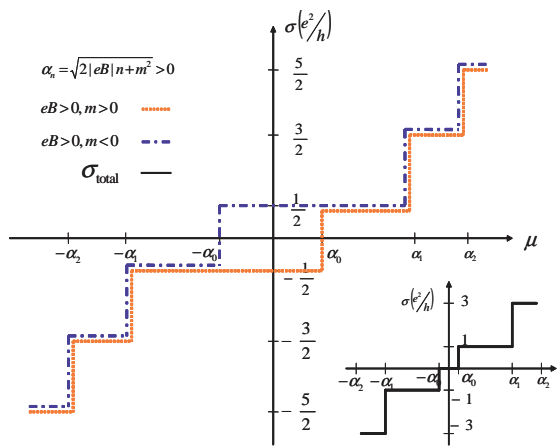}
\caption{Hall conductance as a function of the chemical potential.
The red and blue are the individual conductances from the two Dirac
cones whereas the yellow one is the total Hall conductance. }
\label{diraccond}
\end{figure}

This analysis is correct for the fermion located around the
$K$-point, but as mentioned before the graphene bandstructure
contains two such fermions. For the purpose of being well defined,
we consider a small positive Semenoff mass $m$ at $K$ which means a
small negative mass at $K'$. Consider the case of $eB>0$. The Hall
conductance gets a contribution from both fermions and is zero when
the Fermi level is in the gap $ -m < \mu <m $ and \emph{odd} integer
otherwise. This is then an \emph{odd integer quantum Hall effect} as
in Fig[\ref{diraccond}]. When the gap is vanishingly small, $m
\rightarrow 0$ the region of zero Hall conductance becomes
infinitely narrow.

\section{Harper Equation For Graphene}
 We now present a different argument that reproduces the experimental results and is valid for both high and low $B.$
 The solution to this problem is to carefully
examine the band structure and edge states of graphene in a
magnetic field with rational flux $\phi=p/q.$  The analysis is
based on a generalization of Hatsugai's work\cite{Hatsugai1993a}
to the honey-comb lattice.  The energies of the bands and edge
states are found as zeroes of certain polynomial equations. By
using general polynomial theory we are able to characterize the
bands, find the number of band crossings, and determine the
conditions for zero modes and edge states. By identifying the Hall
conductance as the winding index of the edge state around the band
gap, we find that, as the magnetic field is decreased, the winding
number of the edge states starts taking odd-integer values due to
electron bands collapsing in pairs. The theoretical spectrum is
obtained, and in addition, exact diagonalization results are
presented to support it. It will be evident from the calculated
bandstructure that for large magnetic fields the Dirac argument
does not apply because the Hall conductances of bands at
low-filling do not form a sequence of odd-integers in this case,
as predicted by the relativistic argument.

We use the Landau gauge $A_y = B x$, $A_x=0$, with $B=2 \Phi/ 3
\sqrt{3} a^2$, where $\Phi= p/q$ is the flux per plaquette (hexagon)
and $p, q$ are relatively prime integers. With a Peierls
substitution the effect of the magnetic field is $c_i^\dagger
c_j\rightarrow c_i^\dagger c_j \exp{\int_{j}^i \vec{A} d\vec{r}}$.
In this gauge $k_y=k$ is a good quantum number and the Hamiltonian
for each $k$ is:
\begin{equation}
 H(k) = -t \sum_j c_{k,2j-1}^\dagger c_{k,2j} A_j(k) + c_{k,2j}^\dagger c_{k,2j+1} +
h.c.
\end{equation}
\noindent where
\begin{equation}
A_j(k)=e^{i  \pi \frac{p}{q} (j - \frac{5}{6})} + e^{-i  \pi
\frac{p}{q} (j - \frac{5}{6})}e^{i k}
\end{equation}
\noindent Note that we have not included a mass term in our
tight-binding Hamiltonian because graphene is essentially massless.
Since $A_{j+q} = (-1)^p A_j$ the Hamiltonian is periodic with period
$2 q$, ($A_{j+ 2q} = A_j$) but the energy spectrum, which depends
only on $|A_j|$ is periodic with period $q$. We start with the
one-particle states $|\Psi(k,\phi)\rangle
=\sum_{i}\psi_i(k,\phi)c^\dag_{k,i}|0\rangle$ and act on these with the
Hamiltonian to obtain the equation $H|\Psi\rangle=E|\Psi\rangle.$
There are two independent amplitude equations, one for $i$ odd and
one for $i$ even:
\begin{eqnarray}
& \epsilon \psi_{2j-1} + A_j \psi_{2j} + \psi_{2j-2} =0 \nonumber \\
& A_j^\star \psi_{2j-1} + \epsilon \psi_{2j} + \psi_{2j+1} =0
\end{eqnarray}
\noindent where $\epsilon =E/t$ with $E$ the energy. There are now
two Harper equations for the hexagonal lattice, in contrast to the
single Harper equation for the square lattice. After some
manipulation we find
in a transfer matrix formalism:
\begin{equation}
\left(%
\begin{array}{c}
  \psi_{2j+1} \\
  \psi_{2j} \\
\end{array}%
\right) = \frac{1}{A_j} \tilde{M}_j \left(%
\begin{array}{c}
  \psi_{2j-1} \\
  \psi_{2j-2} \\
\end{array}%
\right)
\end{equation}
with
\begin{equation}
\tilde{M}_j = \left(%
\begin{array}{cc}
  \epsilon^2 - A_j A_j^\star  & \quad\quad \epsilon \\ ~ \\
  -\epsilon & -1 \\
\end{array}%
\right).
\end{equation}
\noindent As opposed to the transfer matrix for the square lattice,
which hops by one site and is linear in energy\cite{Hatsugai1993a},
the graphene transfer matrix hops by two sites and is quadratic in
energy. This reflects the lattice periodicity. Since
$\tilde{M}_{j+q} = \tilde{M}_j$, the periodicity of the energy
spectrum is $q$. We can now define the transfer matrix over the
magnetic unit cell:
\begin{equation}
M(\epsilon)  = \left(%
\begin{array}{cc}
  M_{11}(\epsilon) & M_{12}(\epsilon) \\
  M_{21}(\epsilon) & M_{22}(\epsilon) \\
\end{array}%
\right)\equiv \tilde{M}_{q} \tilde{M}_{q-1}... \tilde{M}_1.
\end{equation}
\noindent By induction we find that $M_{11} $ is a polynomial of
order $(\epsilon^2)^q$, $M_{12}$ and $M_{21}$ are of the form
 $\epsilon\times(\epsilon^2)^{q-1},$  while $M_{22}$ is a polynomial of
order  $(\epsilon^2)^{q-1}$. These polynomials have coefficients
which depend on $k$ and the magnetic flux. We pick our sample of
order $L_y = 2 q l$, commensurate with the magnetic unit cell, where
$l$ is a large integer and the factor of $2 $ is added because we
will require periodic conditions $\psi_{L_y} = \psi_0$, hence $L_y
\equiv 0 \equiv even$. The transfer matrix across the length of this
sample is $M^l$. From Hatsugai\cite{Hatsugai1993a} we know that the
important polynomial to consider is:\begin{equation}
[M^l]_{21}(\epsilon) =0\label{fundamentalpoly}
\end{equation}\noindent The entire spectrum of energy levels for each $k$ value comes from
the zeroes of this polynomial of which there are $L_x -1$. Some of
these states are bulk states and others are edge states. We will now
characterize the edge and bulk states (bands).

It is easy to find one solution to Eq.[\ref{fundamentalpoly}].
Simply take $M_{21}(\epsilon)=0$ and this will imply that
Eq.[\ref{fundamentalpoly}]is satisfied since all upper-triangular
matrices remain so when multiplied by another upper-triangular
matrix. Hatsugai argues\cite{Hatsugai1993a} that the energies of the
\emph{edge states} are given by the zeroes of exactly this
polynomial: $M_{21}(\epsilon) =0$. Since $M_{21}(\epsilon) \sim
\epsilon \times (a(\epsilon^2)^{q-1}+b(\epsilon^2)^{q-2}+\ldots),$
there is always one $\epsilon =0$ solution (zero mode edge state)
which does not disperse and $2(q-1)$ non-zero energy
solutions(edge-states) which come in pairs: $ - \mu_{q-1} \leq
-\mu_{q-2}\leq ...\leq-\mu_1 \leq 0 \leq \mu_1 \leq .... \leq
\mu_{q-2} \leq \mu_{q-1}$. Depending on whether
$M_{11}(\mu_i)/|A_q....A_1|$ is $<$,$>$, or $= 1$ the edge state
will be localized on the left edge, right edge, or be degenerate
with the bulk \emph{i.e.} touching a bulk state\cite{Hatsugai1993a}.

\begin{figure}[h]
\includegraphics[scale=0.37]{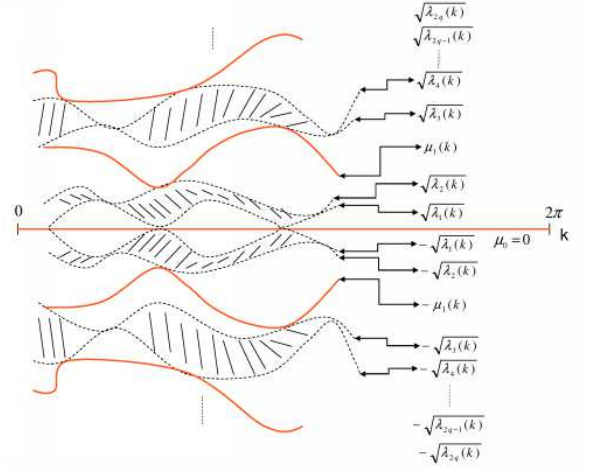}
\caption{Schematic plot of the bulk band structure and edge states obtained from the transfer matrix
formalism. Edge states are solid lines while bulk bands are denoted by the shaded areas bounded by dash and dotted lines.} \label{eigenvalues}
\end{figure}

The bulk states are obtained from the lattice periodicity $j
\rightarrow j+q$ and the Bloch condition:
\begin{eqnarray}
& \left(%
\begin{array}{c}
  \psi_{2q+1} \\
  \psi_{2q} \\
\end{array}%
\right) = \rho(\epsilon) \left(%
\begin{array}{c}
  \psi_{1} \\
  \psi_{0} \\
\end{array}%
\right)
\end{eqnarray}
with  $\rho(\epsilon)$ a pure imaginary phase, {\it i.e.},
$|\rho(\epsilon)|=1$. We also note that we have the transfer matrix
equation
\begin{equation}
\left(%
\begin{array}{c}
  \psi_{2q+1} \\
  \psi_{2q} \\
\end{array}%
\right) = \frac{1}{A_q A_{q-1}...A_1} M \left(%
\begin{array}{c}
  \psi_{1} \\
  \psi_{0} \\
\end{array}%
\right).
\end{equation}
Therefore, combining these two, $\rho(\epsilon)$ is an eigenvalue of
the $2\times 2$ transfer matrix
\begin{equation}
\rho^\pm = \frac{1}{2 A_q .. A_1} [Tr M \pm
\sqrt{(Tr M)^2 - 4 |A_q ....A_1|^2}].
\end{equation}
where we have used $Det[M] = Det[M_q]...Det[M_1] = |A_q...A_1|^2$.
It is easy to see that the Bloch condition $ |\rho(\epsilon)|^2=1$ is satisfied
for $(Tr M)^2 - 4 |A_q ... A_1|^2<0$, based on the fact that
$\rho^+\rho^-=1$. Since $M_{11}$ and
$M_{22}$ are both polynomials of order $q$ in $\epsilon^2$ the
solutions are again paired. Let us rewrite
\begin{equation}
(Tr M(\epsilon^2))^2 - 4
|A_q ... A_1|^2 = \prod_{i=1}^{2q} (\epsilon^2 - \lambda_i),
\end{equation}
with
$0<\lambda_1\leq\lambda_2\leq\cdots\leq\lambda_{2q}$. The energy bands are thus
\begin{equation}
\begin{cases}
\lambda_{2j+1} \le \epsilon^2 \le \lambda_{2j+2} & \quad  \text{bulk state}  \\ ~ \\
\lambda_{2j} \le \epsilon^2 \le \lambda_{2j+1} & \quad  \text{gap region}
\end{cases}
\end{equation}
for $j=0,1,\cdots,q-1$ and $\lambda_0=0$. The edge states lie in the gap region of
the bulk band structure and the $\mu$'s are given by
\begin{equation}
\mu_j \in [\lambda_{2j}, \lambda_{2j+1}]   \quad  j=1,\cdots,q-1.
\end{equation}
We hence have $2q$ energy bands bounded by $4q$
$\lambda$'s, there are $2q-1$ gaps and $2q-1$ edge states as in
Fig[\ref{eigenvalues}].

Besides the above results of band structure, many more details about
the spectrum can already be learned from the behavior of
the function $|A_q A_{q-1} ... A_1|(k)$,
\begin{enumerate}
\item The Hall conductance can be determined from the number of $k$'s that
satisfy

$M_{11}(\mu_{[j/2]}) M_{22}(\mu_{[j/2]}) =|A_q A_{q-1} ... A_1|^2(k_1)$.
\item The first bulk eigenvalue touches the zero energy edge state at the $k$ points where $|A_q ... A_1|(k)=1$.
\item Bulk band width vanishes at $k$ if $|A_q ... A_1|(k)=0$.
\end{enumerate}

For graphene, $|A_q A_{q-1} ... A_1|(k)$ can be explicitly written
as
\begin{eqnarray}
 |A_q A_{q-1} ... A_1|(k) =2^q \prod_{j=1}^q  \left| \cos\left(\frac{k}{2}
+ \frac{(5-6j) p \pi}{6q} \right)\right|.
\end{eqnarray}
\noindent The periodicity of this function is $2\pi/q$.
Hence the number of $k$'s at which $|A_q A_{q-1} ...
A_1|(k)=0$ is equal to $q$ while the number of $k$'s at which $|A_q
A_{q-1} ... A_1|(k)=1$ is equal to $2q$.

We shall now show how to obtain the details of the band structure
from $|A_q A_{q-1} ... A_1|(k)$. Let us assume that the edge state
$\mu_{[j/2]}(k)$ touches the bulk at some point $k=k_1$
\begin{equation}
\mu_{[j/2]}(k_1)= \epsilon_j(k_1)=\pm\sqrt{\lambda_j(k_1)}
\end{equation}
where $[j/2]$ represents the largest integer less than or equal to
$j/2$ and $j= 1,..., 2q$. Since $\epsilon_j(k_1)$ is on the band
edge, we have
\begin{equation}
M_{11}(\mu_{[j/2]})+M_{22}(\mu_{[j/2]})= \pm 2 |A_q A_{q-1} ... A_1|(k_1). \label{edgetouchband1}
\end{equation}
 From the edge state condition $M_{21}(\mu_{[j/2]}(k_1)) =0$, we also know
\begin{equation}
 M_{11}(\mu_{[j/2]}) M_{22}(\mu_{[j/2]}) =
|A_q A_{q-1} ... A_1|^2(k_1). \label{edgetouchband2}
\end{equation}
We hence have
\begin{eqnarray}
M_{11}(\mu_{[j/2]}(k_1))
=M_{22}(\mu_{[j/2]}(k_1)) = \pm |A_q... A_1|(k_1)
\end{eqnarray}
when the edge state touches the bulk state. Thus, we can determine
how many times the edge state starts from $\lambda_{2[j/2]}$ at
$k_1$ and goes up in energy to touch $\lambda_{2[j/2]+1}$ at some
$k_2$ then comes down again to touch $\lambda_{2[j/2]}$ at some
$k_3$ \emph{etc}. This defines the number of wrappings around the
gap and represents the Hall
conductance\cite{Hatsugai1993a,Hatsugai1993b}.

As a function of the momentum $k$ the first bulk eigenvalue
$\lambda_1$ might touch the zero energy line (the zero mode) when
$\epsilon_1(k) =0$. This happens when
\begin{eqnarray}
M_{11}(0) &=& (-1)^q |A_q A_{q-1} ... A_1|^2; \\
  M_{22}(0)&=&(-1)^q; \\
M_{21}(0) &=&0.
\end{eqnarray}
 But from the previous analysis we know that when a
bulk state touches an edge state $M_{22} = \pm |A_q... A_1|$. Hence
the first bulk eigenvalue touches the zero energy edge state in $2q$
points in the first Brillouin zone, namely where $|A_q... A_1|=1$.
This result is confirmed by our exact diagonalization, which will be presented
later.

Using polynomial theory we can in fact prove a more stringent
constraint. We separate the polynomial of order $2q$:
$(Tr
M(\epsilon^2))^2 - 4 |A_q ... A_1|^2= (Tr M(\epsilon^2) - 2|A_q ...
A_1|)(Tr M(\epsilon^2) + 2|A_q ... A_1|).
$
 Now, denote the
eigenvalues of the two subfactors as $g$ for ``green" and $b$ for
``blue" respectively(for purpose of making the connection with the
plots) and put them in ascending order:
\begin{eqnarray}
& Tr M(\epsilon^2) - 2|A_q ... A_1| = \prod_{j=1}^q(\epsilon^2 -
\lambda_j^{g}),
\nonumber \\ & Tr M(\epsilon^2) + 2|A_q ... A_1| =
\prod_{j=1}^q(\epsilon^2 - \lambda_j^{b}),\nonumber
\end{eqnarray}
where $\lambda_1^g < \lambda_2^g< ...< \lambda_q^g$ and $\lambda_1^b
< \lambda_2^b< ...< \lambda_q^b$. \noindent (For $q=2$ the $<$ sign
changes into $\le$ due to the fact that in that case the system
doesn't break T and we can have gapless states.) Depending on
whether $q$ is even or odd we have the following order:
\begin{eqnarray}
& q \; odd: \; \lambda_1^b \le \lambda_1^g < \lambda_2^g \le
\lambda_2^b < ...< \lambda_q^b \le \lambda_q^g  \nonumber \\  &  q
\; even: \; \lambda_1^g \le \lambda_1^b < \lambda_2^b \le
\lambda_2^g < ...< \lambda_q^b \le \lambda_q^g
\end{eqnarray}
\noindent We can see that bulk states are between  ``blue" and
``green" eigenvalues $|[\lambda^b_i, \lambda^g_i]|,$ whereas the
gaps are in between the consecutive ``blue"-``blue" and
``green"-``green" eigenvalues $|[\lambda^g_i, \lambda^g_{i+1}]|$ or
$|[\lambda^b_i, \lambda^b_{i+1}]|$. As such, the width of a band is
$|\sqrt{\lambda^g_i} - \sqrt{\lambda^b_i}|$. The band will become
infinitely thin when $\lambda^g_i =\lambda^b_i,$ or when
$|A_q...A_1|=0$. This happens at $q$ points in the first Brillouin
zone.

As an example, we can see everything above explicitly for the case
$q=3, p=1$. There are $2q-1=5$ edge states with energies $-\mu_2(k), - \mu_1(k), 0, \mu_1(k), \mu_2(k)$, where
\begin{widetext}
\begin{eqnarray}
\mu_{1,2}(k) = \sqrt{3+ \cos{(k - \frac{7 \pi}{9})} + \cos{(k -
\frac{ \pi}{9})} \mp \sqrt{\frac{5}{2} -\cos{(2k - \frac{8
\pi}{9})}+ \cos{(k - \frac{7 \pi}{9})}(2+ \cos{(k - \frac{7
\pi}{9})}) + (\cos{(k - \frac{7 \pi}{9})})^2}}\nonumber
\end{eqnarray}
\end{widetext}
\noindent There are $q=3$ points in the Brillouin zone where each
band becomes infinitely thin given by
\begin{equation}
|A_3 A_2 A_1|(k) = 0\;\;  at \;\; k= \frac{4 \pi}{9}, \frac{10
\pi}{9}, \frac{16 \pi}{9}.
\end{equation}
\noindent  The bands closest to zero energy touch the zero energy
mode at $2q=6$ places in the Brillouin zone where
\begin{equation}
|A_3 A_2 A_1|(k) = 1\;\;  at \;\; k= \frac{\pi}{3}, \frac{5 \pi}{9},
\pi, \frac{11 \pi}{9}, \frac{5 \pi}{3}, \frac{17 \pi}{9}.
\end{equation}
\noindent We also find that the condition $M_{11} (\mu_1(k)) =
M_{22} (\mu_1(k))$ is satisfied at two points in the Brillouin zone,
which means that, in the first gap, the edge state touches the lower
band $\lambda_1^g$ once and the upper band $\lambda^g_2$ also once,
hence the Hall conductance is one. This is the same for when the
Fermi level rests in the second gap, the condition $M_{11}
(\mu_1(k)) = M_{22} (\mu_1(k))$ being  satisfied for two points in
the Brillouin zone as well (see Fig[\ref{bandstructureQ3}]).


\section{Hall Conductance in Graphene}
This section contains the theoretical results from the transfer
matrix approach, as illustrated in the previous section, and the
numerical results from exact diagonalization.  The Hall conductance
in graphene is defined, as usual, as the number of times the edge
state wraps around the gap between neighboring energy bands. The
number of left or right edge states that traverse the entire way
across the gap is the Hall conductance.  We then look at the
evolution of the bands and edge states as the magnetic field is
varied from very strong to weak. We will see how the edge states and
band configuration for strong magnetic field, which do not match
experiment, evolve into the weak-field limit, which does match the
experiments.

This is accomplished in two ways: first, the ``theoretical" edge
states and band structure are found by numerically solving for the
zeroes of the characteristic polynomials $M_{21}(\mu_i) =0$ and
$(TrM (\lambda^{g,b}_i)) \pm 2|A_q...A_1|=0$ introduced in the
previous section. We plot only the $\epsilon \ge 0$ states, the
negative energy states being a mirror image. We also confirm the
theoretical picture by exact numerical diagonalization of the
Hamiltonian matrix for a relatively large number of lattice sites.

\begin{figure}[h]
\includegraphics[scale=0.4]{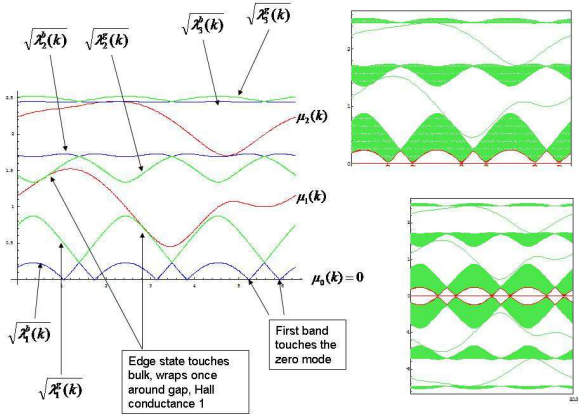}
\caption{Left: Theoretical edge state and band structure
configuration. The edge states are in red while the bulk bands are
in between the consecutive blue and green lines ($\lambda^b,
\lambda^g$). The number of times an edge state wraps around the bulk
is the Hall conductance, which in this case is unity $1$ for the
both the first gap and the second gap. Right: The band structure
obtained from direct diagonalization, upper right just the
$\epsilon>0$ lower right is the full spectrum which is just a mirror
image of the $\epsilon>0$ spectrum. From now on, we will plot only
the positive energy part.} \label{bandstructureQ3}
\end{figure}

We start with $q=3$ in Fig[\ref{bandstructureQ3}].  We see that the
Hall conductance is unity for a Fermi level in either the first or
second gap, clearly in contradiction with the Dirac argument which
would give $\sigma_H=1$ or $3$ depending on which gap. The number of
bands is $2q=6$, there are $2q-1=5$ gaps and edge states, $q=3$
spots where each band becomes infinitely thin, and $2q=6$ points
where the first band touches the zero energy mode.

\begin{figure}[t]
\includegraphics[scale=0.44]{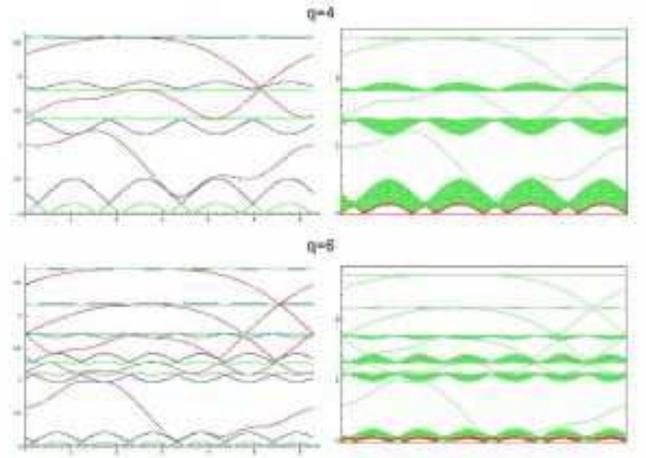}
\caption{Left: Theoretical edge state and band structure
configuration for $q=4$ and $q=6$. Right: Direct diagonalization.}
\label{bandstructureQ4and6}
\end{figure}

We now continue by decreasing the magnetic field to $q=4$ and then
$q=6$, see Fig[\ref{bandstructureQ4and6}]. For $q=4$ we have
 $\sigma_{xy} = 1$ for a Fermi level in the first gap, $\sigma_{xy}
 =2$ for Fermi level in second gap, and $\sigma_{xy} = 1 $ for Fermi
 level in third gap. For $q=6$ the sequence is $\sigma_{xy}
 =1,4,3,2,1$ from the first to the fifth gap.  This does not match the experimental observation of
$\sigma_{xy}= 1,3,5,7, etc.$

One crucial observation to notice is that, as we increase $q$
(decrease the magnetic field), the second and third bulk bands
become closer and closer together in energy; the gap between them
becomes smaller and smaller over the whole Brillouin zone.
Eventually the second and third bands move entirely together upon
increasing q (lowering B). For $q>12$, one cannot distinguish
between the second and third band (nor can one distinguish the edge
states between these bands). The second and third band have
``collapsed" into a new band, a process which we call ``band
collapse."  After these bands have collapsed there are
distinguishable gaps between the first band and the combined band,
and then between the combined band and the fourth band. There are
edge states between the top of the combined band and the fourth
band, and these give $\sigma_{xy} = 3$, for the Fermi level in what
is now the second gap. If we then go to the next gap, this again
does not match the experiment, with Hall conductance being $8.$

\begin{figure}[h]
\includegraphics[scale=0.44]{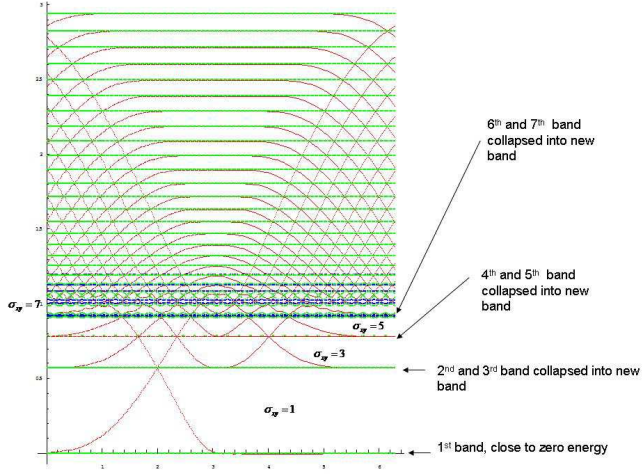}
\caption{Left: Theoretical edge state and band structure
configuration for $q=31$.} \label{bandstructureq31}
\end{figure}

By increasing $q$ even further, we see that the fourth and fifth
bands collapse in a similar fashion, and the gap between them
vanishes uniformly across the $k$ spectrum as they become a single
new band. This happens around $q=22$. The edge states between the
collapsed second and third bands and the collapsed fourth and fifth
bands remain the same as before, giving $\sigma_{xy}=3$ but now the
edge states between the collapsed fourth and fifth bands and the
sixth band give $\sigma_{xy}=5$. This process repeats itself while
$q$ is increased. The total number of bands increases when $q$ is
increased. But some of these bands collapse together so that we
cannot distinguish them unless we have infinite resolution. We
present the results for $q=31$ Fig[\ref{bandstructureq31}].

Upon increasing $q$ the band collapse leads to double degeneracy of
each of the bands \emph{except} the zero energy band, and this gives
the odd integer Hall conductance in graphene. This is beautifully
seen as the number of positive or negative-slope edge states that
disperse in the resolvable gaps. Hence the experimental situation is
theoretically confirmed as the weak-field limit of graphene.

\begin{figure}[t]
\includegraphics[scale=0.44]{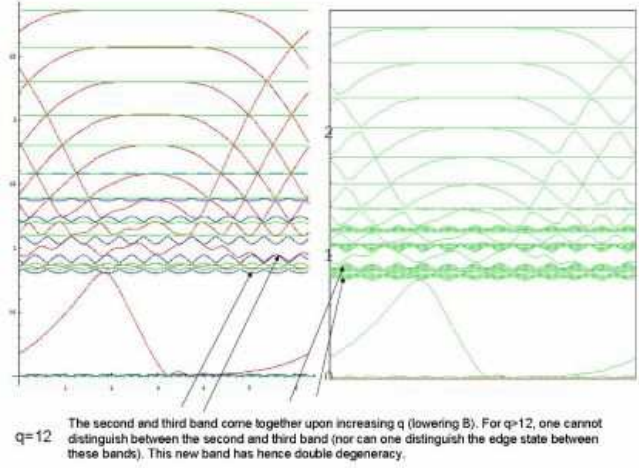}
\caption{Left: Theoretical edge state and band structure
configuration for $q=12$. Right: Direct diagonalization.}
\label{bandstructureq12}
\end{figure}

\begin{figure}[h]
\includegraphics[scale=0.44]{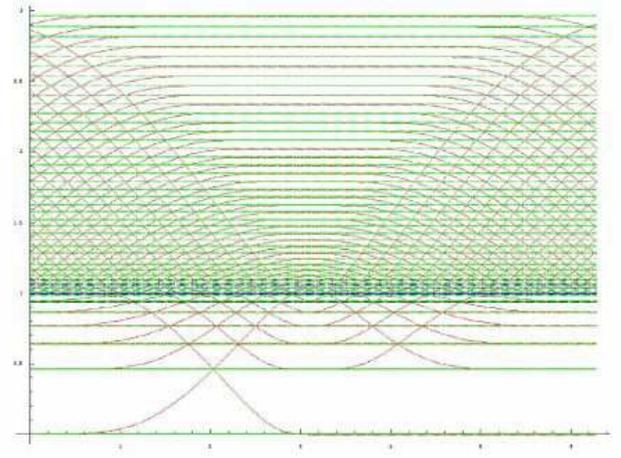}
\caption{Left: Theoretical edge state and band structure
configuration for $q=49$.} \label{bandstructureq49}
\end{figure}

The theoretical band structure can actually be continued to large
$q$, as the polynomials are well behaved.  We give the $q=49$ plot
as well, where we can see all the odd-integer quantum hall effects
from $1,3,5,7,9,11,$ see Fig[\ref{bandstructureq49}].

By examining the common properties of each bandstructure plot it
appears that the spectrum of bands and edge-states can be classified
into two parts: a relativistic section and a non-relativistic
section. This structure originates from the original tight-binding
dispersion relations without the B field, \bea \epsilon(k_x,k_y)=\pm
t\sqrt{1+ 4\cos^2 \frac{\sqrt{3} }{2} k_x +4 \cos \frac{\sqrt{3} }
{2} k_x \cos \frac{3}{2} k_y}.
\nonumber \\
\eea The Dirac nodes are located at $(\pm\frac{4\pi}{3\sqrt{3}},0)$
and $(\pm \frac{2\pi}{3\sqrt{3}}, \pm \frac{2\pi}{3})$. The
linearized dispersion relations persist up to around $E\approx t$.
Above this energy scale the bands become parabolic. Accordingly, in
Fig. \ref{bandstructureq49}, $\sigma_{xy}=1,3,5$ ... at low energy
and the energy of bulk levels goes as $E_n\approx\sqrt{n},$ a
feature of relativistic Landau levels. On the other hand
$\sigma_{xy}=1,2,3,....$ starting from the top of the bands (where
parabolic bandstructure is expected) and there is almost equal
spacing between each of these Landau levels, which is a feature of
the harmonic-oscillator-like non-relativistic Landau levels. A
$\sigma_{xy}$ of $1$ is seen in the first gap from the band ceiling
and increases by one for each Landau level below the top. A similar
thing occurs for the non-relativistic levels near the bottom of the
set of bands. The crossover region is at $E\approx t,$ where the
band collapse occurs.

The odd-integer sequence shown in Fig[\ref{bandstructureq49}] is
clearly represented in the experimental data which, as stated
before, is in the low magnetic field limit of graphene. With a
flux $\phi= 1/q$ in each unit cell the magnetic field is $\sim
\frac{1.3\times 10^5}{q} \;\; \rm{Tesla}$ which is a very large
magnetic field. For experimentally realizable magnetic fields we
would expect $q\sim 1000$ and the odd-integer sequence would be
continued to larger values. Abnormalities in this sequence would
not arise until more Landau levels were filled. Overall, there
will be a sequence (possibly very long) of odd-integer quantum
Hall conductances followed by conductances which do not follow a
certain pattern. Then there will be a
relativistic-non-relativistic crossover region where the Landau
level spacings change character from $n^{1/2}$ to $n$. The
non-relativistic energy levels will then persist to higher
energies.
\subsection{Effect of Disorder}
\begin{figure}[t]
\includegraphics[scale=0.44]{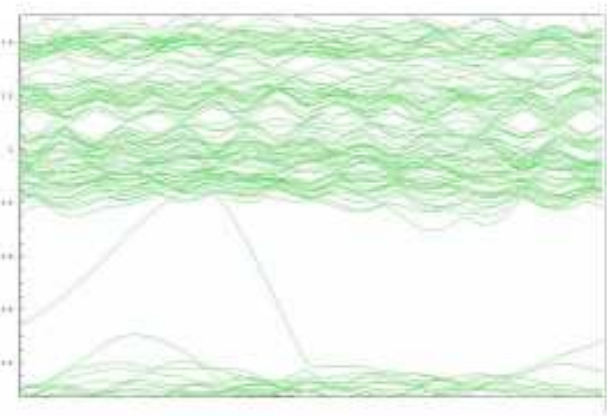}
\caption{Numerical calculation for edge states and band structure
for $q=10$ and disorder variance $0.15.$} \label{disorder1}
\end{figure}

\begin{figure}[h]
\includegraphics[scale=0.44]{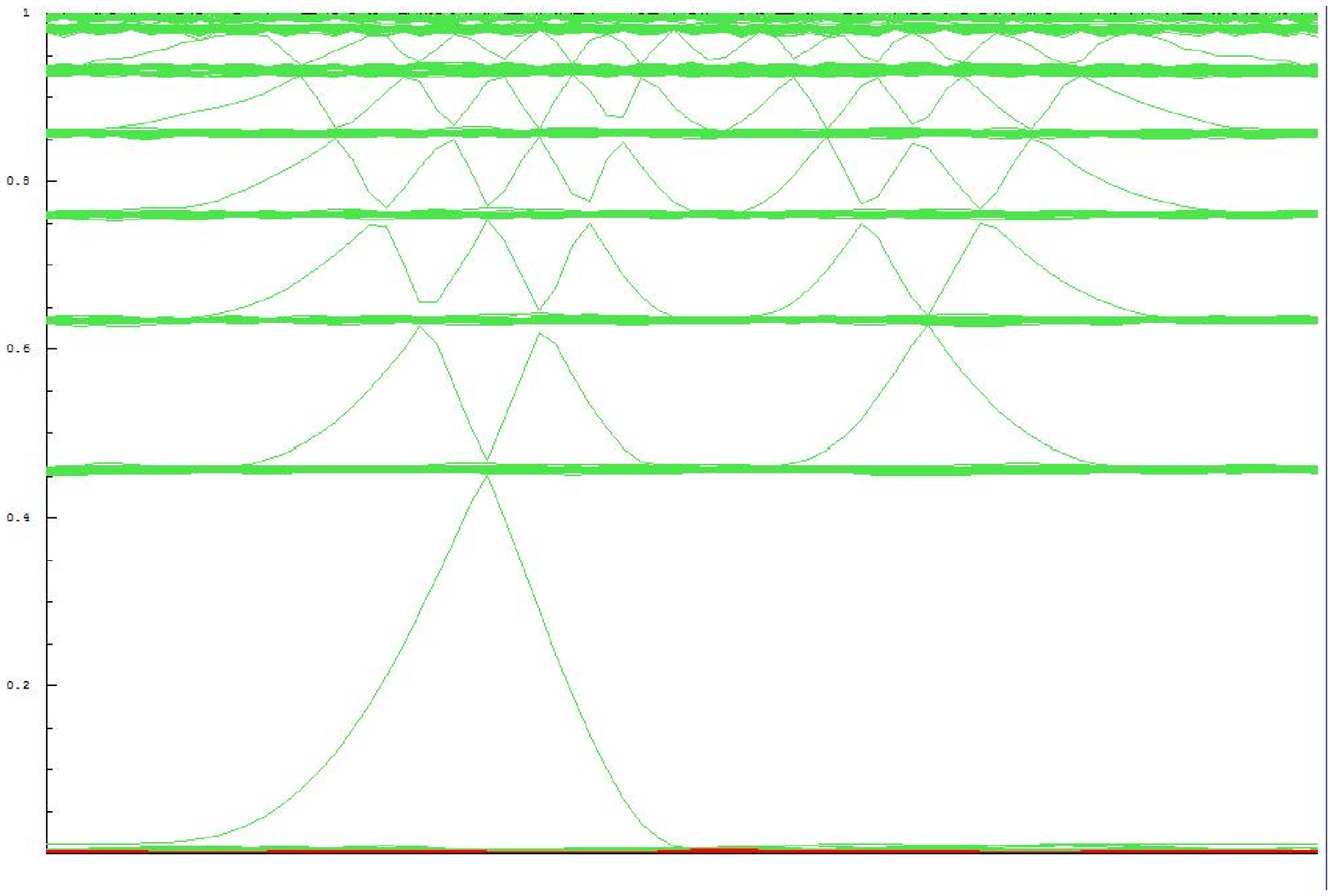}
\caption{Numerical calculation for edge states and band structure
for $q=50$ and disorder variance $0.01.$} \label{disorder2}
\end{figure}

 We have considered the stability of the edge-states under disorder. Although not
tractable analytically, we were able to use
 numerical diagonalization (which up to now has remarkably
matched the analytic results) to study the introduction of disorder
into the system. The disorder term we added to the system is:
\begin{equation}
H_{dis}(k)=X_{D} \delta_{IJ}
\end{equation}\noindent where $X_D$ is a random variable with
gaussian distribution and mean $0,$ and $I,J=1,2,\ldots L-2,L-1.$
We also tested a uniform distribution for the $X_D$ with
essentially the same results. For relatively high disorder
\emph{e.g.} the variance of $X_D\sim 0.15$ the structure of the
lowest energy edge state is robust (see Fig[\ref{disorder1}]).
However, the edge states representing higher plateaus, such as
$n=3,5,7,9,11,\ldots$ are washed out. Note that the hopping
parameter is defined to be $1,$ so $0.15$ is very high disorder.
For lower disorder, with the variance of $X_D\sim 0.01,$ all of
the edge states are clearly visible up to the
relativistic-non-relativistic crossover (see
Fig[\ref{disorder2}]), just as in the disorder-free plots given
above; \emph{e.g.} as in Fig[\ref{bandstructureq12}].

\subsection{Non-zero Semenoff term}
\begin{figure}[!ht]
\includegraphics[scale=0.40]{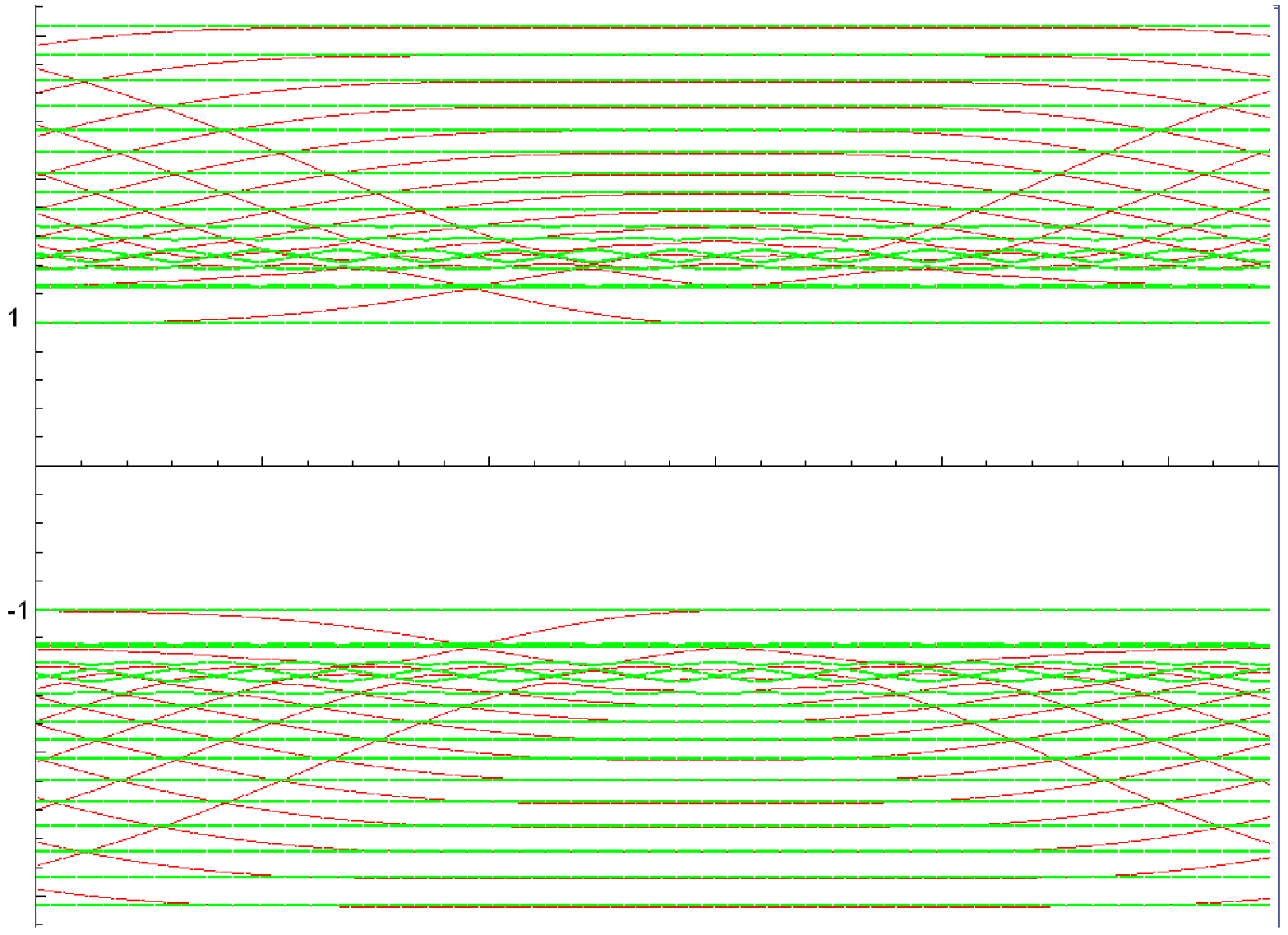}
\caption{(color online) (a)Theoretical edge state and band
structure configuration for $q=17$, with Semenoff mass $m=1.$}
\label{bandstructureq17mgap1}
  \end{figure}
The previous formalism can be easily extended to incorporate the
case of a non-zero Semenoff mass $m$. As an example, for Boron
Nitride (BN) the hamiltonian has the form:
\begin{eqnarray}
 H = -t \sum_j ((c_{2j-1}^\dagger c_{2j} A_j + c_{2j}^\dagger
 c_{2j+1}+h.c.)\nonumber \\
 + m(c_{2j-1}^\dagger c_{2j-1} - c_{2j}^\dagger c_{2j}))
\end{eqnarray}
\noindent The new Harper's equations are:
\begin{eqnarray}
& (\epsilon +m) \psi_{2j-1} + A_j \psi_{2j} + \psi_{2j-2} =0 \nonumber \\
& A_j^\star \psi_{2j-1} + (\epsilon -m) \psi_{2j} + \psi_{2j+1} =0
\end{eqnarray}
\noindent and the transfer matrix now becomes:
\begin{equation}
\tilde{M}_j = \left(%
\begin{array}{cc}
  \epsilon^2- m^2 - A_j A_j^\star  & \epsilon -m  \\
  -(\epsilon+m) & -1 \\
\end{array}%
\right)
\end{equation}
\noindent As we can see, $\tilde{M}_q \tilde{M}_{q-1}
....\tilde{M}_1 = (\epsilon+m) \times
 P^{(q-1)}(\epsilon^2)$ where $P^{(q-1)}(\epsilon^2)$ is a
 polynomial of order $q-1$ in $\epsilon^2$. Hence the former zero
 energy edge state has now moved to $\mu_0 = -m$. There are no edge
 states between $[-|m|, +|m|]$, but the rest of the analysis
 applies. We plot the band structure for $m=1$, $q=17$ (see Fig[\ref{bandstructureq17mgap1}]).

\section{Spin and Valley Splitting in the $n=0$ Landau Level}

\begin{figure}[!ht]
$\begin{array}{c}
   \subfigure{
          \label{beta0}
\includegraphics[scale=0.34]{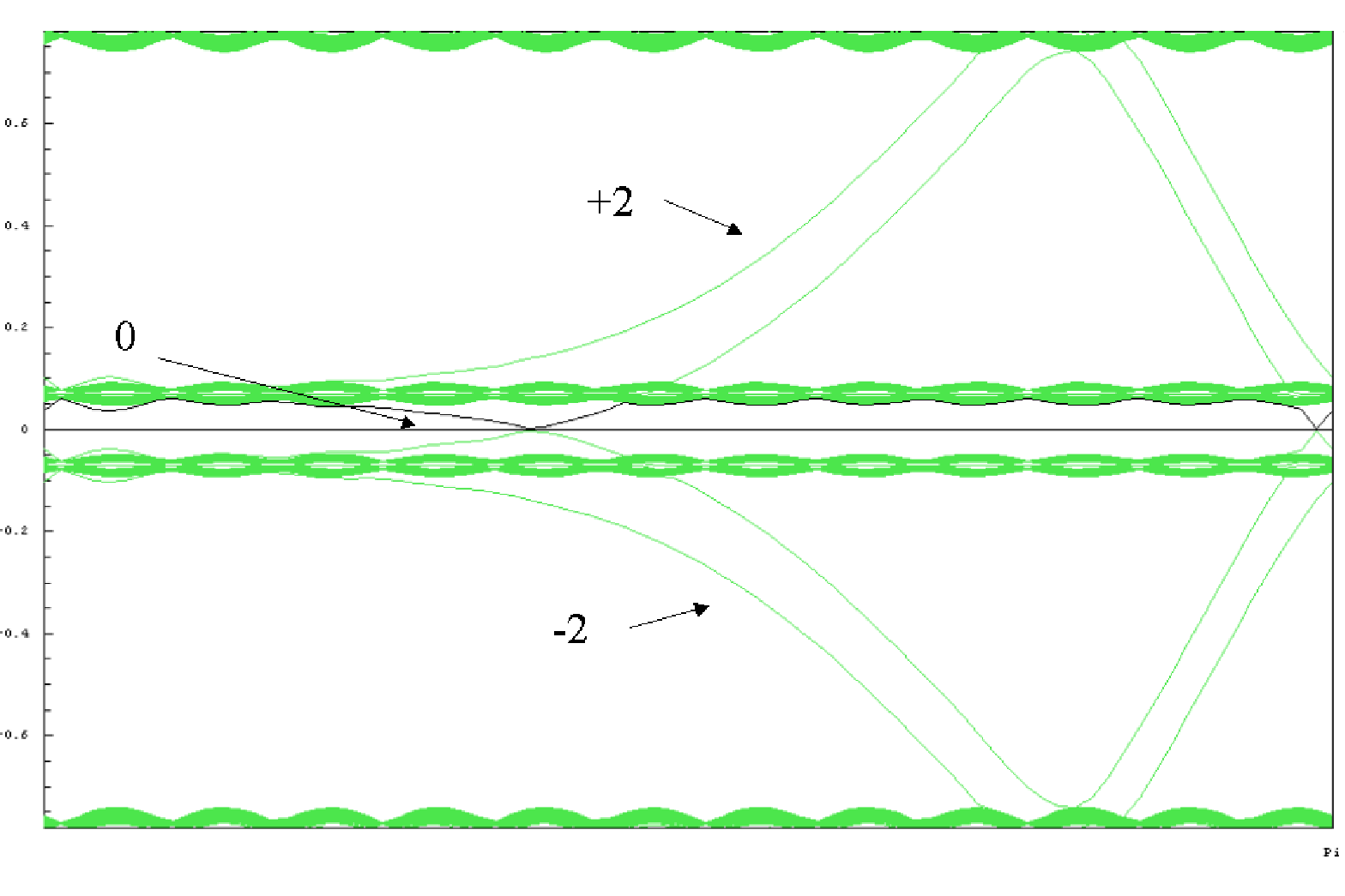}} \\
 \subfigure{
          \label{beta50}
\includegraphics[scale=0.35]{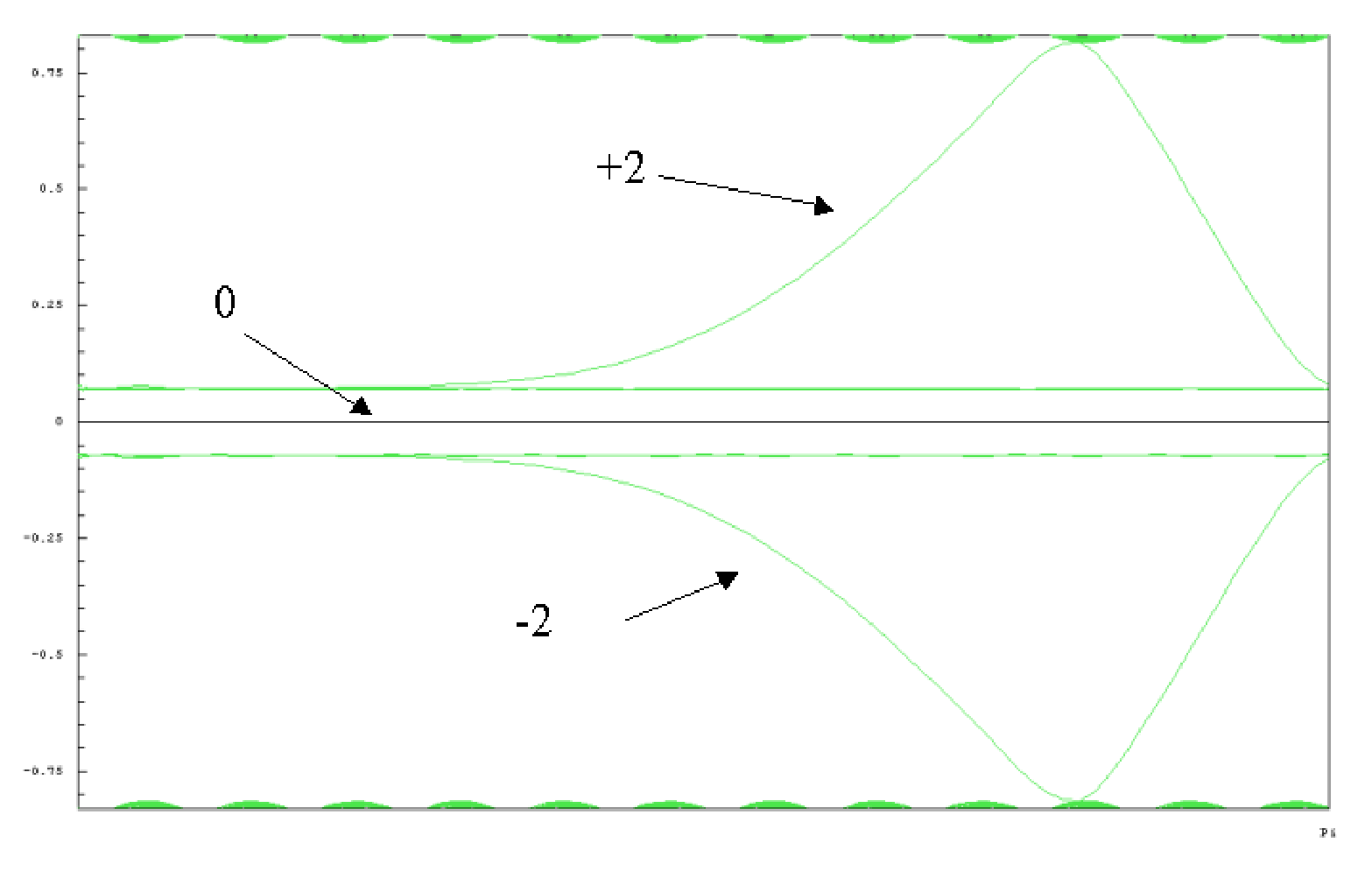}}\\
\subfigure{
          \label{beta5}
\includegraphics[scale=0.35]{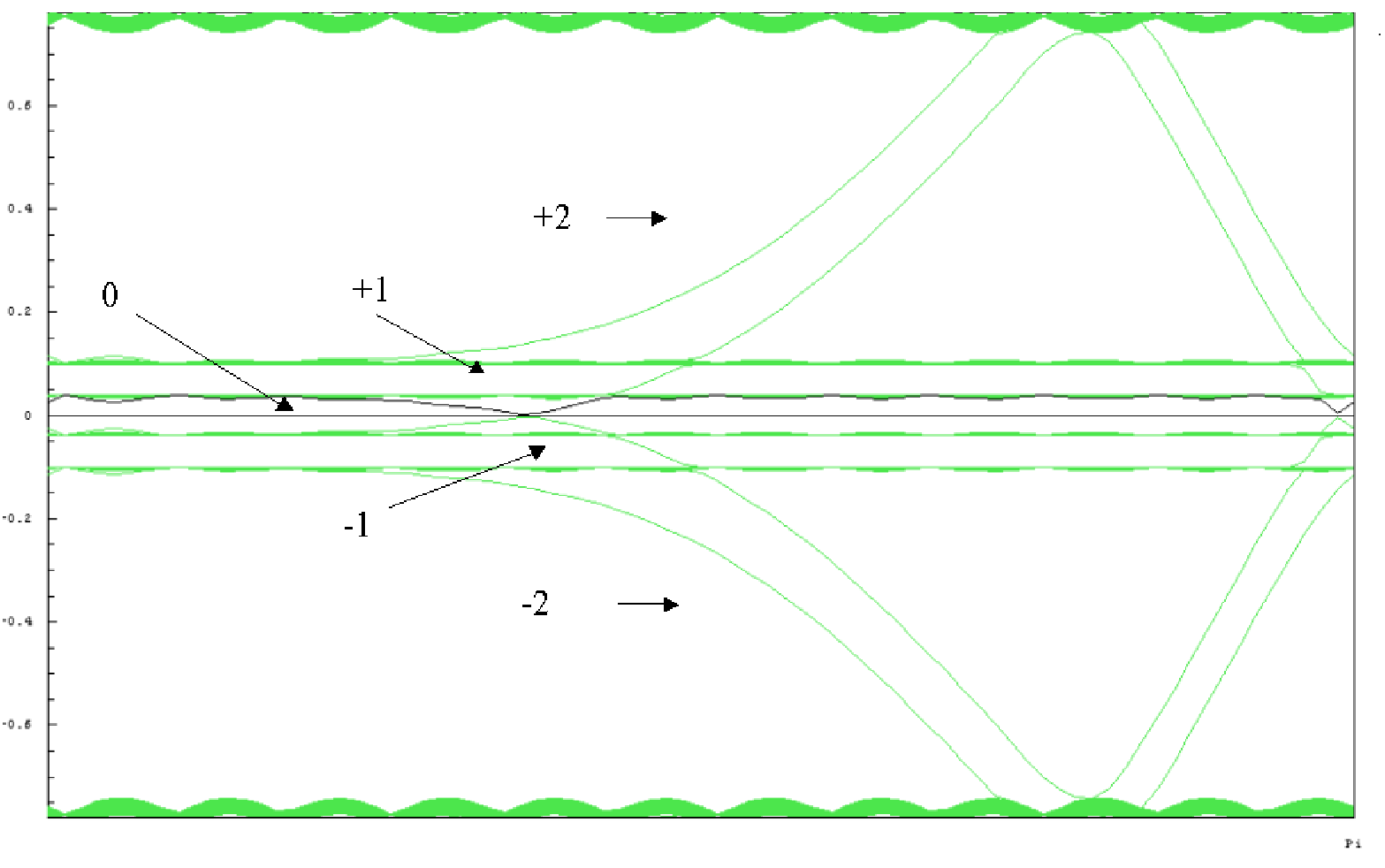}}
\end{array}$
\caption{Landau level bands and edge states around E=0 for (a)Spin
splitting only (b)Valley splitting only (c) Spin and valley
splitting. Quantum Hall conductances for particular gaps are noted
by the integer labels and are in units of $e^2/h.$}
  \end{figure}

We now focus on the breaking of spin and/or valley degeneracy in
the $n=0$ Landau Level. The idea of spin splitting is very natural
since $g\sim 2$ in graphene and there is a large magnetic field
applied perpendicular to the sample. Splitting the valleys
however, is more subtle since there is no natural alternating
sublattice potential or applied strain. We investigate the changes
to the QHE plateau structure and edge states when these splittings
can be resolved energetically.

First we consider the case of only spin splitting. Due to the
Zeeman effect the spin states in each Landau level will be split
by $g\mu_B B.$ For the $n=0$ Landau level one spin state is pushed
above zero energy and the other is pushed below zero energy. When
the chemical potential lies in the gap at zero energy between the
split spin states there is an additional QH plateau with
$\sigma_{xy}=0.$ The picture is not quite this simple because this
gap, unlike the Semenoff mass gap discussed above, contains edge
states which can be seen in Fig[\ref{beta0}]\cite{albanin2006a}.
Usually the presence of edge states in the gap signals a non-zero
QH conductance but here there is actually one electron edge state
and one hole edge state. These two edge states combine together to
give zero Hall conductance but produce a non-zero spin-Hall
conductivity since they are spin-polarized in opposite directions:

\begin{equation}
\sigma_{spin} = 2 \frac{e^2}{\hbar}.
\end{equation}
\noindent This spin current can be observed in a $4$-terminal
geometry or in a system with magnetic leads.

The case where only the valleys are split in the $n=0$ level, no
matter by what means, is very similar to the case of a non-zero
Semenoff mass given above. As in that case there is a gap at zero
energy leading to an additional zero conductance plateau, however
here there are no edge states in the gap, thus no spin Hall
conductivity. The band picture and sequence of quantum Hall
conductances can be seen in Fig[\ref{beta50}].

Finally we come to the case where there are both spin and valley
splittings. Gaps will appear when the $n=0$ level is unfilled,
$1/4$-filled,$1/2$-filled, $3/4$-filled, and completely filled
yielding a sequence of QH conductances $\sigma_{xy}=-2,-1,0,1,2$
in units of $e^2/h$ when the chemical potential lies in each of
these gaps. The band picture with each of these conductances can
be seen in Fig[\ref{beta5}]. This sequence matches the data
recently produced by \cite{kim2006a} in very high magnetic
fields.In the graphene sample there will be some small valley
splitting due to imperfections (shear strain, impurities, or
surface roughness) but not enough to produce a gap large enough to
exhibit the quantum Hall effect. Since there is no applied strain
we must look for many-body effects that would give rise to this
splitting.

The idea of exchange ferromagnetism, which applies in the
non-relativistic quantum Hall effect is also applicable here with
some differences. In a normal QH system we expect that for the
lowest Landau level we should see valley-polarized ground
states\cite{rasolt1985,rasolt1986}. The $\sqrt{n}$ dependence in
the graphene Landau level spectrum should not be important as long
as the Landau gap is large \emph{i.e.} $\hbar
\omega_c>>e^2/\ell_B,\hbar/\tau.$ The excitation energy of
skyrmions has been calculated in \cite{kallin1984}. However, if we
considered higher Landau levels there would be some quantitative
corrections.

The second thing to consider is the correlation between the valley
index and the sublattice index. For the $n=0$ level if an electron
is in a particular valley then its spatial wavefunction resides on
a single sublattice, A or B. If this Landau level is $1/4$-filled
or $3/4$-filled there will be a valley and spin polarized ground
state. The spin polarization is from the Zeeman splitting and the
system will form a valley-polarized ``ferromagnet"-like state due
to exchange correlations. In this level the valley and sublattice
are correlated, but they are correlated such that if the electrons
reside in only one valley then they reside on a single sublattice
which minimizes the Coulomb interaction. This leads to a
spin-polarized charge modulation where there will be an excess of
charge on one sublattice. This will form a weak charge density
wave with charge density modulation where the percentage of charge
modulation is proportional to $N_0/N_T$, the amount of electrons
in the $n=0$ Landau level divided by the total number of electrons
in the system. The electrons that participate in the charge
modulation are effectively the difference between the number of
electrons at half-filling and the number of electrons currently in
the system.

This valley polarized ground state will produce an interaction gap
characterized by the energy to produce a charged excitation. Since
there is no applied strain we expect that $SU(2)$ valley skyrmions
will be cheaper to create than particle-hole excitations
\cite{sondhi1993}. We do not expect to see full $SU(4)$ skyrmions
because the $g$-factor in graphene is not small\cite{sondhi1993}.
This raises the possibility of measuring valley skyrmions in
graphene as was recently done in AlGaAs \cite{shkolnikov2005}.
Since we are projecting into the $n=0$ Landau level we can use the
calculation of \cite{kallin1984,arovas1999} to estimate the spin
stiffness and thus give an estimate of the energy to create a
skyrmion:
$E_{sk}=4\pi\rho_s=\frac{1}{4}\sqrt{\pi/2}(e^2/\epsilon\ell_B).$
If we compare the energy width of the plateau of the spin-split
states to that of the valley-split states shown in \cite{kim2006a}
they are roughly of the same order of magnitude. However
$E_{sk}/(g\mu_B B)\sim 54$ at $B= 45 T$ so our skyrmion energy is
clearly an overestimate. For the valley skyrmions measured in
AlGaAs\cite{shkolnikov2005} the data also clearly shows that
$E_{sk}$ is an overestimate by a factor of $\sim 40$ for their
systems at zero applied strain. This factor compensates for the
overestimation and brings the skyrmion energy to the right order
of magnitude.  Another interesting fact is that at low magnetic
field this valley splitting gap vanishes and the $\sigma_{xy}=\pm
1$ plateaus disappear. This could be the result of there being two
few electrons in the $n=0$ Landau level to produce this well
correlated effect. Overall the valley degeneracy splitting
suggests that small spin-polarized charge density modulation or
valley skyrmions could be measured in graphene.

\section{Conclusion} We have shown that the ``relativistic" quantum
Hall effect in graphene has its origin in a band-collapse picture
where two bands become degenerate upon decreasing the flux per
plaquette. A series of exact results for the honeycomb lattice are
given, as well as an index theorem for the number of dirac modes
in a magnetic field. At large magnetic fields, the system has a
transition between ``relativistic" and non-relativistic QHE. When
the spin-gap is resolved, the system exhibits a spin-Hall effect
due to existence of opposite spin electron and hole edge states in
the gap. We discussed the effects of disorder and adding a
Semenoff mass term. We concluded with discussion on spin and
valley splitting in the $n=0$ Landau level and its implications
for the quantum Hall effect.

\section{Note} During the preparation of this paper, we have noticed
a series of other papers that have independently reached some of the
conclusions presented in this
manuscript\cite{albanin2006a,nomura2006a,fertig2006a,macdonald2006a,hatsugai2006a}.

\section{Acknowledgements}
 B.A.B acknowledges support from the SGF. T.L.H. acknowledges support from
NSF. This work is supported by the NSF under grant numbers
DMR-0342832 and the US department of Energy, Office of Basic Energy
Sciences under contract DE-AC03-76SF00515.

\end{document}